\documentclass[journal,12pt,onecolumn,draftclsnofoot]{IEEEtran}
\ifCLASSINFOpdf
\else
\fi
\usepackage{amsmath,amssymb,amsfonts}
\usepackage{algorithmic}
\usepackage[ruled,vlined,linesnumbered]{algorithm2e}
\SetAlCapNameFnt{\small}
\SetAlCapFnt{\small}

\usepackage{authblk}

\usepackage{graphicx}
\usepackage[justification=centering]{caption}
\usepackage{xcolor}

\begin{document}
\bstctlcite{bstctl:etal, bstctl:nodash, bstctl:simpurl}

%
\title{Multi-Objective Provisioning of Network Slices using Deep Reinforcement Learning}


\author[*,**]{Chien-Cheng Wu}
\author[*]{Vasilis Friderikos}
\author[**]{Čedomir Stefanović}
\affil[*]{Department of Engineering, King’s College London, London, \authorcr United Kingdom} 
\affil[**]{Department of Electronic Engineering, Aalborg University, Copenhagen, Denmark}


%


\maketitle

\begin{abstract}
Network Slicing (NS) is crucial for efficiently enabling divergent network applications in next-generation networks. 
Nonetheless, the complex Quality of Service (QoS) requirements and diverse heterogeneity in network services entail high complexity for Network Slice Provisioning (NSP) optimization.
The legacy optimization methods are challenging to meet various low latency and high-reliability requirements from network applications.
To this end, we model the real-time NSP as an Online Network Slice Provisioning (ONSP) problem.
Specifically, we formulate the ONSP problem as an Multi-Objective Integer Programming Optimization (MOIPO) problem. 
Then, we approximate the solution to the MOIPO problem by applying the Proximal Policy Optimization (PPO) method to the traffic demand prediction. 
Our simulation results show the effectiveness of the proposed method compared to the state-of-the-art methods with a lower SLA violation rate and network operation cost.
\end{abstract}

\begin{IEEEkeywords}
  Network Slicing, Deep Reinforcement Learning, Multi-Objective Optimization, Integer Programming.
\end{IEEEkeywords}

%
\IEEEpeerreviewmaketitle

\section{Introduction}
\IEEEPARstart{N}{etwork} Slicing (NS) is essential in the next-generation mobile wireless networks~\cite{7926920}. 
It enables efficient connectivity to various services with diverse requirements by instantiating multiple logical networks on top of the substrate, i.e., the physical network infrastructure. 
Note that some emerging 5G services, such as those related to the Ultra-Reliable Low Latency Communication (URLLC), require dedicated network resources to achieve the stringent quality of service (QoS) requirements.
NS can offer dedicated network resources for multiple network services mapped and managed over a physical wireless network infrastructure~\cite{8476595}.
In that respect, a network slice can be considered as a self-contained logical network with its physical network resources, topology and traffic flows with established QoS requirements~\cite{8463518}.

In addition, Virtual Network Functions (VNFs) bring higher levels of flexibility as VNFs can be anchored at different network locations and scaled flexibly with NS to meet the fluctuating user traffic demands, 
thus allowing for efficient on-demand Network Slice Provisioning (NSP).
A network slice incorporates a set of VNFs organized in different suitable locations across the transportation paths and depending on the needs of the service.
However, real-time and high-quality resource provisioning for multiple network slices is a formidable task, proved to be NP-hard~\cite{AMALDI2016213}.

In order to address this hurdle, the Cloud-native Network Functions (CNFs) have been actively considered and standardized within the 3rd Generation Partnership Project (3GPP)~\cite{ngmn}.
According to the latest 3GPP architecture~\cite{3gpp.23.501,3gpp.29.531,3gpp.28.531},
We construct slices as CNFs' interconnections in our NS paradigm, which align with the conventional notion of VNFs.

\begin{figure}
  \centering
  \includegraphics[scale=0.40]{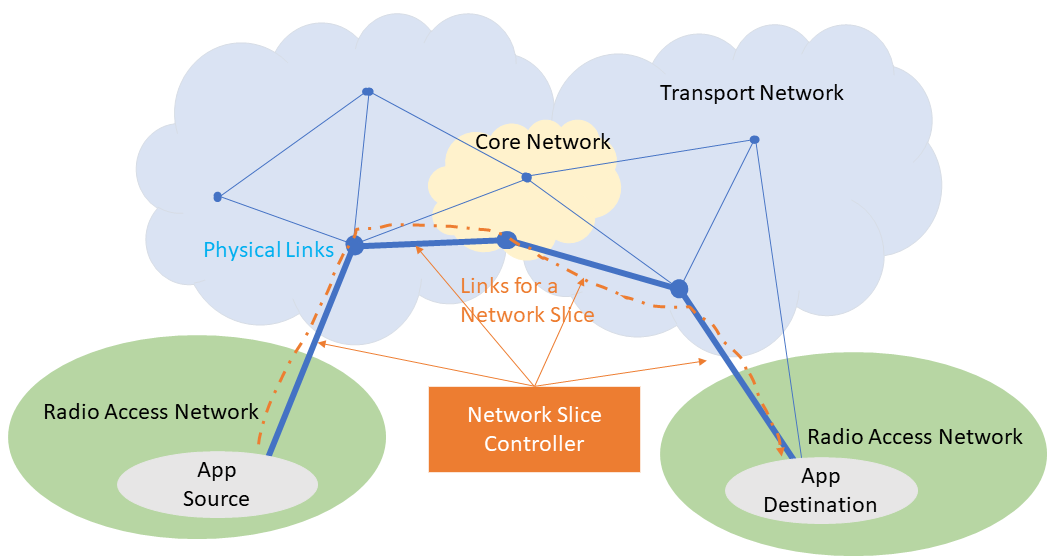}
  \caption{Routes from Source Node to Destination Node}
  \label{fig:architecture}
\end{figure}

Within the paradigm, a slice instance should be constructed by related CNFs' interconnections for a user traffic demand request.
We further separate the CNFs' interconnections into the data rate and delay arrangements over the physical network infrastructure.
Fig.~\ref{fig:architecture} presents how the network slice controller constructed a network slice via associating network resources with a suitable route.
Without loss of generality, a network service provider can append more network resource variables other than the data rates and delays to the paradigm.

This paper investigates efficient slice provisioning as the CNFs' network resource provisioning in the NS paradigm. 
We focus on the data rate and delay arrangements for the slice provisioning.
Our contributions include formulating a multi-objective optimization model and subsequent deep reinforcement learning framework to optimize the network slices provision 
whilst being robust to unknown traffic demand fluctuation of the users. 
The robustness in provisioning is achieved by providing a probabilistic guarantee that the amount of provisioned slices will meet the slice QoS requirements.
We implement the deep reinforcement learning framework by using Proximal Policy Optimization (PPO) algorithm~\cite{ppo} with two neural networks as a value function and a policy function pair.
Under this framework, the network slice controller determines its slice provisioning by the optimal QoS-dependent policy from the PPO algorithm. 
The problem is then solved by approximately the optimal solution with the joined slices allocation and demands prediction.
To illustrate the distinction of the computational performance, we implement a suboptimal approach as a performance benchmark where slice provisioning demands are considered in a batch form, i.e., as several groups sequentially. 
Both solutions are compared to a nominal provisioning scheme that only processes the current user demand without considering relations between the slice provisioning demands.
Through extensive simulations, 
we show that our proposed algorithm can reduce the network operation cost while keeping a low Service-Level Agreement (SLA) violation rate.

The rest of this article is structured as follows. Section~\ref{related_work} analyzes related work and highlights our contributions.
The mobile network model, user demand requests and the network slices are presented in Section~\ref{system_moddel}.
The online NSP problem with uncertainties in the number of users and the associated traffic demands is then formulated in Section~\ref{problem_formulation}.
Next, we present the existing approaches and our reinforcement learning method to solve the robust NSP problem in Section~\ref{methods}. 
Section~\ref{simulation} presents numerical results. Finally, Section~\ref{conclusion} concludes our perspectives.

\section{Related Work}
\label{related_work}
Software-Defined Network (SDN) uses programmable controllers to provide flexible and cost-effective network services in next-generation networks.
NSP is a crucial function of SDN for supporting dynamic user demands in different network services.
The main challenge for the NSP is to optimize the deployment of slices across the physical infrastructures matching network services constraints. 
Therefore, we review the existing NSP design and organize the related solutions to the NSP optimization problem.

A slice is realized as a concatenation of communication (wired-cum wireless links) and computing resources that can span across the Radio Access Network, Transport Network, or Core Network~\cite{7892961}.
In the context of rule-based solutions, the authors in~\cite{9495099} propose a practical NS implementation.
The proposed model provides an efficient solution by analyzing historical NS information but the network slices assigned to the same tenant cannot overlap in time.
The research in~\cite{8116371} considers the slice provision with VNF placement for the SDN-based 5G mobile-edge cloud. 
Their algorithm provides a flexible slice provision by placing VNFs in distributed data centers.
Similarly, the work in \cite{7116162} describes a VNF placement algorithm with emphasis on the mobile core network, exploiting the cost of placement to allocate the VNFs. 
Their problem formulation takes physical network constraints into account for different network service capacity and connectivity.

On the other hand, the deep reinforcement learning solutions for dynamic demand optimization continuously attract the attention of researchers.
The authors of~\cite{9137719} propose a novel framework that uses assured resources which are offered based on the forecast of the user demands.
According to the framework, the designed stochastic algorithm can handle the trade-off between the traffic uncertainty in the forecast and overbooking.
The work~\cite{8057090} designed a hybrid machine learning model for spatiotemporal prediction, where an autoencoder models the spatial dependence, 
and the temporal dependence is captured by a Long Short Term Memory neural network.
We also mention~\cite{10.1007/s10922-022-09654-8} that proposes a heuristically-controlled A3C algorithm to demonstrate network slice provision with dynamic traffic. 

To the best of our knowledge, this work presents the first attempt to use Proximal Policy Optimization (PPO) algorithm in multi-objective network slice provisioning.
More specifically, we formulate the network slice provisioning problem as a multi-objective integer programming optimization problem.
Then we propose a deep reinforcement learning framework that allows for online provisioning, which jointly considers the fairness of slice provision and the cost of network operation.

\section{System Model}
\label{system_moddel}
We assume the system is time-slotted. 
Each action, such as time increment or user demand request arrival, happens at the beginning of a slot and completes before the end of that slot.

\subsection{Network Model}
The network is modeled as an undirected graph $\mathcal{G} = (\mathcal{V}, \mathcal{L})$, where $\mathcal{V}$ denotes the set of network nodes and $\mathcal{L}$ is the set of undirected wired-cum-wireless links that constitute the deployed physical infrastructure in the network.
By $l_{i,j} \in \mathcal{L}$ we denote a link between nodes $v_{i}$ and $v_{j}$ ($v_{i}, v_{j}\in \mathcal{V}$).
The nodes in the graph represent either a radio access network component, e.g., a base-station (BS), a transportation regional network component or a core network element. 
We assume that all network nodes are virtualized, which means that they are capable of running virtual machines and/or containers over the corresponding physical (bare metal) machine capabilities.
Furthermore, we denote the available link capacity as a vector $C=[ C_{1,1}, \dots, C_{i,j}, \dots, C_{V,V}]$, 
where each element $C_{i,j} \geq 0$ is the capacity of direct link  $l_{i,j}$  between nodes $v_{i}$ and $v_{j}$;   
if there is no link between $v_{i}$ and $v_{j}$, then the value of $C_{i,j}$ is zero.
We also denote the occupied link capacity as a function 
\begin{equation}
 0 \leq \alpha(i,j) \leq C_{i,j}.
\end{equation}
The delay of the links 
is given in the vector $D=[D_{1,1}, \dots, D_{i,j}, \dots, D_{V,V}]$, 
where each element $D_{i,j} \geq 0$ express the delay between $v_{i}$, and $v_{j}$ 
The active network operation cost of links in $\mathcal{L}$ is given by the vector $P=[P_{1,1}, \dots, P_{i,j}, \dots, P_{V,V}]$, 
where each element $P_{i,j} \geq 0$ is the active operation cost between nodes $v_{i}$ and $v_{j}$.\footnote{If there is no link between $v_{i}$ and $v_{j}$, then we assign the maximum available integer value to $D_{i,j}$ and $ P_{i,j}$ in our simulations to represent the disconnected scenario.}

\subsection{User Demand Model}
We assume that the user demand requests at a time slot $t$ is a set, denoted by $\mathcal{R}(t)$. 
Each request $r$, $r \in \mathcal{R}(t)$, comes from a network application of an user, asking for a network slice template at time slot $t$.
All requests are generated according to a Poisson arrival model determined by a aggregate arrival rate $\lambda$ (requests/slot).
A user demand request is represented as a tuple $r=\{v_{s}, v_{u}, b_{r}, d_{r}, \mathcal{T}_{r}, a_{r}, h_{r}\}$. 
where $v_{s}$ and $v_{u}$ denote the source and destination nodes respectively, $b_{r}$ is the required data rate, $d_{r}$ denotes the delay requirement, and the demand type is expressed as $\mathcal{T}_r$.
Further, the initial time slot of the demand is $a_{r}$ and the demand life time is $h_{r}$ slots. 
Thus, the demand will start from the time slot $a_{r}$, continue for $h_{r}$ slots, and becomes terminated at the time slot $t=a_{r} + h_{r}$.
We assume there are finite number of demand types and network nodes. 
When the network slice controller receives a request, 
the controller decides whether to accept or reject it in an online manner.

For example, at a time slot $t = t_{1}$, request set $\mathcal{R}(t_{1})$ is handled by the slice controller and then the appropriate network slices are constructed for the requests.
If a request $r \in \mathcal{R}(t_{1})$ cannot be served at $t_{1}$, it will be stored in a queue until the time reached the time slot ($t=a_{r}+ h_{r}$).
Thus, at any time slot $t$ there are a set of requests, $\mathcal{R}(t)$, that includes both new arrived requests as well the backlogged requests 
(i.e. the requests that are in the queue, which currently cannot be served and whose termination time has not yet expired).

\subsection{Network Slice Model}
Each slice $n$ can be denoted as
\begin{align}
n = \{ v_{n,s,r}, v_{n,u,r}, E_{n,r}, y_{n,r}, \mathcal{T}_{n,r} \}
\end{align}
where $s$ is the index of source node $v_{n,s,r}$, $u$ is the index of destination node $v_{n,u,r}$; $(v_{n,s,r}, v_{n,u,r} \in \mathcal{V})$.
Further, $E_{n,r}$ is a set of virtual links $E_{n,r}=\{e_{i,j}^{n,r}\}$ 
with defined provisioning resources.
Each virtual link $e_{i,j}^{n,r} \in E_{n,r}$ can be mapped to one and only one physical link $l_{i,j}$ between node $v_{i}$ and node $v_{j}$, and $E_{n,r}$ is a loop-free path, starting from $v_{n,s,r}$ and ending at $v_{n,u,r}$.
The loop-free path represents the connectivity constraints of the slice $n$.
The capacity and delay constraints of a virtual link $e_{i,j}^{n,r}$ are the same as the ones of the mapped physical link $l_{i,j}$.
The slice load is $y_{n,r}$, which means each $e_{i,j}^{n,r} \in E_{n,r}$ occupies $y_{n,r}$ capacity of link $l_{i,j}$;
therefore $e_{i,j}^{n,r}= y_{n,r}$ 
$y_{n,r} \leq \min_{\forall e_{i,j}^{n,r} \in E_{n,r}}  \alpha_{i,j}$.
The type of slice is denoted as $\mathcal{T}_{n,r}$ which is allocated according to the demand type.
Finally, the slice delay is $\sum_{\substack{\forall e_{i,j}^{n,r} \in E_{n,r}}} {D_{i,j}}$, where $i, j$ are in $e_{i,j}^{n,r} \in E_{n,r}$.
A feasible slice can only be provided by its slice controller by fulfilling the connectivity, capacity and delay constraints.
\begin{figure}
  \centering
  \includegraphics[scale=0.60]{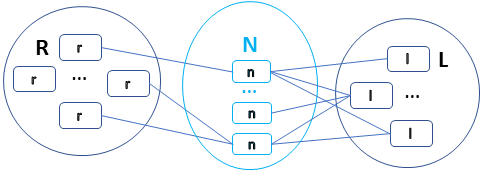}
  \caption{Network Slices Mapping Demonstration}
  \label{fig:formulation}
\end{figure}

\section{Problem Formulation}
\label{problem_formulation}

Our goal is to generate network slices that accommodate the underlying user requests in terms of data-rate and latency whilst maximizing the deployment fairness and minimizing the deployment cost, formally defined later in the section.
In particular, we avoid network congestion at some links by maximizing the deployment fairness.
To this end, we fairly distribute requests to the whole network.
We also consider the co-existence of multiple slices on the same physical network
and assume that each slice serves network traffic from a single source to a single destination, 
and therefore we represent a slice by a source-destination pair.

For each user traffic demand $r$, the NS controller aims to generate an appropriate network slice $n$ to serve $r$.
For example, in Fig.~\ref{fig:formulation}, there is user request set $R$ and link set $\mathcal{L}$. 
The NS controller aims to construct a set of network slices $N$ to serve $R$ using (a part of) link set $\mathcal{L}$.
A request $r$, $\forall r \in R$, can only be served by a single slice $n$, $n \in N$, and each $n$ consists of a subset of $\mathcal{L}$.
A virtual link $e_{i,j}^{n,r}$ comes from slice $n$ and request $r$. Moreover, the virtual link $e_{i,j}^{n,r}$ is made by the physical link $l_{i,j}$ between node $v_{i}$ and node $v_{j}$ in $\mathcal{L}$.

The slice generation procedure can be decomposed into two steps. 
The first step is to find potential end-to-end routes which can be utilized for request $r$ from its source node $v_{n,s,r}$ to its destination node $v_{n,u,r}$. 
The second step is to find a route which can minimize the objective function while satisfying the constraints.
To represent the link utilization, we define a utilization matrix $X$: for each matrix element, $x_{i,j}^{n,r}$, 
if slice $n$ from request $r$ utilizes link $l_{i,j}$, then $x_{i,j}^{n,r}=1$, otherwise, $x_{i,j}^{n,r}=0$.
Since multiple slices share the same physical network,
the bandwidth allocation at each virtual link shall not exceed the available physical link bandwidth.
We formulate the global network constraint as
\begin{equation}
  \label{global_bandwidth_constraint}
  \sum_{\substack{\forall n \in N \\ \forall r \in \mathcal{R}(t) }} x_{i,j}^{n,r} \cdot e_{i,j}^{n,r}  \leq C_{i,j}.
\end{equation}

In addition, each virtual route in slice $n$ mapped onto the physical links should satisfy the requested user traffic volume
\begin{equation}
  \label{local_bandwidth_constraint}
  \min_{\substack{\forall e_{i,j}^{n,r} \in  E_{n,r}}} \alpha(i,j)  \geq b_{r}.
\end{equation}

Moreover, each virtual route in slice $n$  
should also satisfy the requested user traffic latency
\begin{equation}
  \label{route_latency_constraint}
  \sum_{\substack{\forall n \in N \\ \forall r \in \mathcal{R}(t) }} x_{i,j}^{n,r} \cdot D_{i,j}  \leq d_{r}.
\end{equation}

From the above constraints and matrices, we formulate the Online Network Slicing Provisioning (ONSP) problem as:
\emph{Given the user demand requests, $\mathcal{R}(t)$, how to generate a set of slices $N$ with minimum cost and maximum fairness to transmit the user traffic?}


Specifically, the cost objective function is defined as the summation of all the slices' cost in Eq.~(\ref{cost_objective_function}) below.
\begin{equation}
  \label{cost_objective_function}
  f_{1} = \min {\sum_{\substack{\forall n \in N \\ \forall l_{i,j} \in \mathcal{L} }}} x_{i,j}^{n,r} \cdot P_{i,j}
\end{equation}
The cost optimization strategy tends to prioritize the low-cost virtual links during slice provisioning.
However, the priority of low-cost links should be limited within a propor level.
Therefore, we refer~\cite{jain1984index} to define the fairness objective function as the ratio of allocated data rate and available data rate Eq.~(\ref{fair_objective_function}) below.
\begin{equation}
  \label{fair_objective_function}
  f_{2} = \max {\dfrac
                {\left(\sum_{\substack{\forall n \in N \\ \forall l_{i,j} \in \mathcal{L}}}\dfrac{\alpha(i,j)}{C_{i,j}}\right)^{2}}
                {\lvert N \rvert {\sum_{\substack{\forall n \in N \\ \forall l_{i,j} \in \mathcal{L}}} \left(\dfrac{\alpha(i,j)}{C_{i,j}}\right)^{2}}}.
              }
\end{equation}

The ONSP problem is then reformulated as a Multi-Objective Integer Programming Optimization (MOIPO) problem to construct a route for a slice, 
and Breadth-First Search (BFS) search algorithm is used to find the minimal cost paths through the network.
In general, at a timeslot $t$, given a set of user demand requests, $\mathcal{R}(t)$ as an input,
our algorithm aims to generate a set of network slices $N$ and keep the related bandwidth and latency guarantees for every network slice.

\section{Purposed Methods}
\label{methods}

\subsection{Greedy Approach}
\label{Greedy}
The greedy algorithm is used as a less computationally intensive benchmark because it only considers one possible provisioning sequence when allocating virtual links.
The computational complexity is $\mathcal{O}(\mathcal{V}+\mathcal{L}+\mathcal{R})$.
Algorithm~\ref{algo:greedy} shows a greedy ONSP algorithm.
An undirected graph, $\mathcal{G}$, and user demand requests set, $\mathcal{R}(t)$, are given as input.
We use the cardinality of a set such as $|\mathcal{R}(t)|$ and $|\Omega|$ to measure the number of elements in the set $|\mathcal{R}(t)|$ and  the set $|\Omega|$ respectively.  
From the graph $\mathcal{G}$ and the set $\mathcal{R}(t)$, 
we can obtain the bandwidth constraint functions and latency constraint functions, see Eq.~\ref{global_bandwidth_constraint},~\ref{local_bandwidth_constraint} and \ref{route_latency_constraint}.
Then, the greedy algorithm iterates all requests in $\mathcal{R}(t)$ by a loop with finite steps.
For each user demand request $r \in \mathcal{R}(t)$, we use the BFS algorithm to search the graph for all possible sets of links 
that satisfy the connectivity constraint of $r$.

A set of links can construct a unique path with a fixed sequence from the BFS algorithm.
If a path can satisfy the connectivity constraint of $r$, we define the path is a feasible path. Otherwise, the path is not feasible.
Next, we consider all feasible paths and check the link availability of each path based on the given bandwidth and latency constraints.
If a link is not available, the corresponding path is skipped. Otherwise, we continue to the next step.

Next, we calculate the cost and fairness values based on the objective functions.
After that, we compare the cost value and fairness value in the current step with the cost value in the previous step.
If the cost value in the current step is less than the cost value in the previous step and the fairness value in the current step is higher than the fairness value in the previous step, 
we assign the path $\omega$ to be a candidate for slice $n$ with respect to the request $r$.
Next, to generate a slice $n$ by a non-empty candidate. Otherwise, move to the next feasible path.
Finally, we aggregate each slice $n$ as a set $\mathcal{N}$ for output.

\begin{algorithm}
  \small
	\SetAlgoLined
	\DontPrintSemicolon
	\KwIn{A undirected graph $\mathcal{G} = (\mathcal{V}, \mathcal{L})$ and user demand requests,  $\mathcal{R}$}
  \KwOut{network slices,  $\mathcal{N}$}
  Sort the set $\mathcal{R}$ by the initial time\\
	\For{$r= 1, 2, 3, \cdots$ until $r= |\mathcal{R}|$}{
		Find all feasible paths, $\Omega$, based on the BFS algorithm\\
    \For{$\omega= 1, 2, 3, \cdots$ until $\omega= |\Omega|$}{
      Check the link availability in $\omega$ based on the constraint functions.\\
      Calculate the cost value and fairness value based on the objective functions.\\
      Compare with the previous cost value and fairness value.\\
      \If{current value is better than previous value}{
        Assign the current $\omega$ to candidate\\
      }
    }
    \If{candidate is not empty}{
      Generate a network slice $n$ by the candidate\\
    }
	}
  Aggregate all network slices to $\mathcal{N}$\\
	\caption{Greedy Algorithm}
	\label{algo:greedy}
\end{algorithm}

\subsection{Integer Programming Approach}
\label{IP}
Integer Programming (IP) was introduced to model a series of optimization problems.
Many alogrithms had tried to efficiently solve the Integer Programming problem such as the Constraint Integer Programming (CIP)~\cite{10.1007/978-3-540-68155-7_4}.
In addition, The CIP method had been proven to be able to obtain the optimal solution in~\cite{10.1007/978-3-540-68155-7_4}.
The general form of CIP is the following: given an finite set $\mathcal{H}$ of constraints $h_{i}: \mathcal{R}^{n} \to \mathcal{Z}$, $i=\{1, \dots, m\}$,
 a variable set $\mathcal{X}$, $x \in \mathcal{Z},\forall x \in \mathcal{X}$, and a vector of objective functions $f \in \mathcal{R}^{n}$,
derive an optimal solution $\theta$ from $\mathcal{X}$ if the CIP is satisfiable as $\theta = \min\{ f^{T}\mathcal{X} | h_{i}(X) = \text{true}, \forall h_{i} \in \mathcal{H}\}$.
Each request at a specific timeslot $t$ in the ONSP problem in Section~\ref{problem_formulation} can be written in the CIP form:
\begin{equation*}
  \begin{array}{ll@{}ll}
  \text{minimize}  & \displaystyle\sum\limits_{k=1}^{2}\sum\limits_{n=1}^{|N|}\sum\limits_{i=1}^{|V|}\sum\limits_{j=1}^{|V|} f_{k} \cdot x_{ij}\\
  \text{subject to}& \displaystyle\sum\limits_{n=1 ,\dots, |N|}   \alpha(i,j) \cdot x_{ij} \leq C_{i,j}       & 
                    \{i,j:e_{ij} \in \mathcal{L}\}\\
                   & \displaystyle\sum\limits_{\forall n \in N}   D_{i,j} \cdot x_{ij} \leq d_{i,j}& \{i,j:e_{ij} \in \mathcal{L}\}\\
                   & x_{i,j} \in \{0,1\}, &i,j=1 ,\dots, |V| \\
  \end{array}
\end{equation*}

The considered optimization problem is indeed an integer programming problem~\cite{10.1007/978-3-540-68155-7_4}.
In each time slot, the slice controller receives a set of user demands as a request instance, $\mathcal{R}(t)$.
The goal of IP algorithm is to construct a set of slices and to satisfy all or part of the requests.
The computational complexity for IP algorithm to search for the optimal solution is equal to $\mathcal{O}((\mathcal{V} + \mathcal{L}) \times \mathcal{R}(t) \times \mathcal{R}(t))$~\cite{complexity} .

\subsection{Deep Reinforcement Learning Approach}
\label{PPO}

Deep Reinforcement Learning (DRL) had been discussed in~\cite{kool2018attention} to solve IP optimization problems.
Hence, we enhance the DRL framework to solve the formulated ONSP problem under the IP model.
The framework aims to construct an agent in the slice controller for the network slices provisioning. 
The agent will not only learn to find an optimal solution at a specific NSP instance, but reuse what it has learned in previous instances.
In such cases, the PPO algorithm has been shown to achieve a higher performance than the other DRL algorithms~\cite{44969}.
This motivates us to embed the PPO algorithm to our framework.

We assume an PPO agent interacting with a network environment to learn a provisioning policy without prior information.
The state space only includes the demand requests and network graph.
Given a set of demand requests $\mathcal{R}(t)$ the provisioning policy $\pi(o_{t}, \theta_{t})$ returns a subset of $\mathcal{R}(t)$ as an action $a_{t}$ to satisfy 
the network constraints in Eq.~(\ref{global_bandwidth_constraint}) and (\ref{local_bandwidth_constraint}) and the traffic demand constraints in Eq.~(\ref{route_latency_constraint}).

The action space consists of a set of actions $a_{t}$ to represent which requests are accepted to transmit in a network slice in slot $t$. 
For example, if only $r_{1}^{t} = 1$ for all $r_{i}^{t} \in \mathcal{R}(t)$, it means that only request $1$ is accepted to obtain network resources and to transmit traffic network slice at a time slot $t$.
The requests for which $r_{i}^{t} = 0$ are rejected to construct their network slices and have to wait for the next action.

Every interaction between the agent and the network environment happens at the beginning of a slot.
In each interaction, the agent samples the environment to get an observation $o_{t}$ and performs an action $a_{t}$ based on the provisioning policy.
After performing the action, the agent receives a reward $R(t)$.
Then the agent waits for the time slot $t+1$ to interact with the network environment.
The learning process repeats the interactions continuously to approximate the optimal provisioning policy, which obtains the maximal reward. 

The reward function is defined as follows.
If the request $r$ is accepted, the reward function is
\begin{equation}
  \label{rej_rwd}
	R(t) = \sum_{\substack{\forall n \in N \\ \forall l_{i,j} \in \mathcal{L}}} x_{i,j}^{n,r} \cdot P_{i,j}, \; \forall r_{i}^{t} = 1 .
\end{equation}
Otherwise, the reward function is
\begin{equation}
  \label{acc_rwd}
  R(t) = \sum_{\substack{\forall n \in N \\ \forall l_{i,j} \in \mathcal{L}}} x_{i,j}^{n,r} \cdot P_{i,j} \cdot (-1), \; \forall r_{i}^{t} \in \mathcal{R}(t) .
\end{equation}

In the beginning, the agent will fetch an initial environment observation $o_0$ from the network environment.
The agent observes a set of metrics in each slot $t$, $o_t$, including the links' capacity, bandwidth and latency requirements of received user demands, and the cost achieved in the last $k$ iterations. 
Then the agent feeds these values to the neural network, which outputs the next action.
The next action is defined by which requests are to be chosen for the next iteration $k+1$, as well as how much bandwidth they will be allocated, at time slot $t+1$.
The provisioning policy is transformed from the action obtained from the trained neural network.
If a request is selected to transmit packets, the required bandwidth will be fully reserved. 
Otherwise, the request waits until it gets allocated.
After the new provisioning is deployed to all requests, a reward is observed and fed back to the agent.
The agent uses the reward information to train and improve its model.
Our implementation of the PPO algorithm in the provisioning problem is detailed in Algorithm~\ref{algo:ppo}.
Starting from the initial parameters, the PPO algorithm optimizes its policy $\pi$ until converging or reaching $K$ iterations.
At each iteration $k$, the PPO agent collects observation of a time slot.
Next, it selects an action with the current policy.
After the agent takes the selected action, the agent obtains a reward based on the reward functions in Eq.~\eqref{rej_rwd} and \eqref{acc_rwd}.

In addition, we formulate a Q-value function, a state value function and an advantage function, used to compute the intermediate values in each iteration.
\begin{align}
	Q_{\pi}(o_{t}, a_{t}) & = \mathbb{E}_{o_{t+1}, a_{t+1}}[ \gamma^{t} R(t)] \\
	V_{\pi}(o_{t}) & = \mathbb{E}_{o_{t+1}, a_{t}}[ \gamma^{t} R(t)] \\
	A_{\pi}(o_{t}, a_{t}) & = Q_{\pi}(o_{t}, a_{t}) - V_{\pi}(o_{t})
\end{align}
where $\gamma$ is a discount factor, $\gamma \in [0,1]$.

Then we construct the surrogate loss on these observations and optimize policy with stochastic gradient descent (SGD) for $e$ epochs and minibatch size $\mathcal{B}$.
The computational complexity of PPO algorithm in testing phase is also $\mathcal{O}(\mathcal{V}+\mathcal{L}+\mathcal{R})$.

\begin{algorithm}
  \small
	\SetAlgoLined
	\DontPrintSemicolon
	\KwIn{An initial policy with parameters $\theta_{0}$ and initial observation $o_{0}$, a set of user demand requests $\mathcal{R}(t)$}
  \KwOut{A set of neural network parameters}
	\For{$k= 1, 2, 3, \cdots$ until $k= K$ or convergence}{
		Fetch user demand requests, $\mathcal{R}(t)$, based on observation $o_{k}$.\\
		Take action to select user demand requests using policy $\pi=\pi(\theta_{k})$.\\
		Compute advantage estimation based on the value function.\\
		Optimize surrogate function $\nabla R(t)$ with respect to $\theta_k$ using $e$ epochs and minibatch size $\mathcal{B}$.\\
		$\theta_k \leftarrow \theta_{k+1}$\\
	}
	\caption{Proximal Policy Optimization Algorithm}
	\label{algo:ppo}
\end{algorithm}

\section{Numerical Investigations}
\label{simulation}

\begin{table}[ht]
  \centering
  \caption{Simulation Parameters}
  \label{tab:paratable}
  \begin{tabular}{ |p{5cm}||p{5cm}|}
      \hline
      \hline
      Parameter Name& Value/Distribution(mean, variance)\\
      \hline
  Number of Nodes $|V|$ & $8$ \\
  Number of Physical Links $|L|$ & $12$ \\
  Capacity of Links & Uniform Distribution (100, 200)\\
  Latency of Links & Uniform Distribution (1, 10)\\
  Cost of Links & Uniform Distribution  (1, 20)\\
  Simulation Duration & $1000$ slots \\
  Request Demand Requirement ($b$) & Normal Distribution (0, 0.1) \\
  Request Latency Requirement ($d$) & Normal Distribution (1, 0.1) \\
  Request Initial Time $a_{r}$ & Normal Distribution (0, 0.1) \\
  Request Life Time $h_{r}$ & Normal Distribution (1, 0.1)\\
  Request Arrival Model & Poisson Process \\
      \hline
 \end{tabular}
\end{table}

\subsection{Simulation Environment}
\label{ssec:simenv}
Our simulation is executed on a desktop PC with an Intel i7 CPU and 8 GB memory.
To construct a general simulation environment, we refer to the model in~\cite{newman2003random} to construct the network graph in our simulation.
Because the MOIPO problem is NP-hard and intractable in a network with large network node sets and edge sets, we verify the proposed method with the other two benchmarks on a reduced network scale.
Some algorithms using high-performance computing clusters have been used to solve large-scale problems, but it is beyond the scope of this paper.
Table~\ref{tab:paratable} lists the environment parameters in our implementation.

\subsection{Simulation Results}
\label{ssec:results}
\begin{figure}
  \centering
  \includegraphics[scale=1]{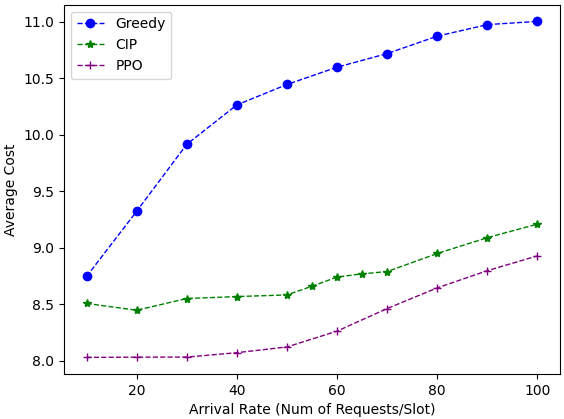}
  \caption{Slices Operation Cost with Different User Request Arrival Rate}
  \label{fig:cost_comparison}
\end{figure}

\begin{figure}
  \centering
  \includegraphics[scale=1]{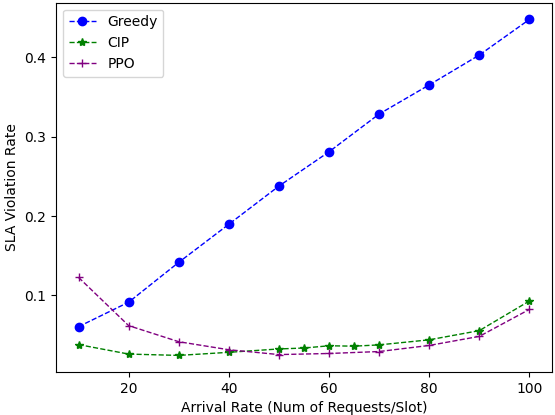}
  \caption{SLA Violation Rate with Different User Request Arrival Rate}
  \label{fig:sla_rate_comparison}
\end{figure}

We validate the performance of the PPO method in terms of the SLA violation rate and network operation cost.
We define the network operation cost is the summation of all the slices' costs with respect to the their physical links and average the cost by all requests. 
In addition, we define the SLA violation rate as the ratio between the number of the provisioning requests and the number of all requests.
We also compare the results of the PPO method with the two benchmarks, Greedy and IP.
Fig.~\ref{fig:cost_comparison} shows the comparison of slice operation cost, and Fig.~\ref{fig:sla_rate_comparison} presents the performance difference of SLA violation rate.
The Greedy algorithm has the worst cost performance and the highest SLA violation rate because the sequential requests processing cannot closely approximate the optimal solution. 
The IP algorithm has better performance in terms of the cost and SLA violation rate as it considers the complete request set in the course of optimization.
On the other hand, we verify the PPO algorithm can learn to approximate the similiar network operation cost and SLA violation rate as the IP algorithm without any prior information used in the IP algorithm. 
From Fig.~\ref{fig:sla_rate_comparison}, the reduction in SLA violation rate compared between CIP and PPO algorithms was ranging from 1.15 to 1.28 times.

\section{Conclusions}
\label{conclusion}
We studied the NSP optimization problem by considering the cost of network operation and the fairness of slice provisioning in a virtualized network. 
To minimize the network operation cost while maximizing the provisioning fairness, 
we implement the PPO algorithm in our DRL framework to predict the incoming user demand and search for the optimal solution to the optimization problem.
We also implemented two benchmark algorithms to demonstrate the performance difference.
Simulation shows the effectiveness of our DRL framework compared with the benchmarks.


%


\bibliographystyle{IEEEtran}
\bibliography{IEEEfull, reference}

\end{document}